\documentclass[english,preprint,JIP]{ipsj}

\usepackage{graphicx}
\usepackage{latexsym}

\usepackage{cite}
\usepackage{amsmath,amssymb,amsfonts}
\usepackage{algorithmic}
\usepackage{textcomp}
\usepackage{xcolor}
\usepackage {booktabs}
\usepackage{url}
\usepackage{enumitem}
\usepackage{placeins}
\def\Underline{\setbox0\hbox\bgroup\let\\\endUnderline}
\def\endUnderline{\vphantom{y}\egroup\smash{\underline{\box0}}\\}
\def\|{\verb|}

\received{2026}{6}{10}
\usepackage[varg]{txfonts}%%!!
\makeatletter%
\input{ot1txtt.fd}
\makeatother%

\begin{document}

\title{Observing the Relationship between QoS Unpredictability,
 Prediction Error, and User Activity in a Remote Desktop Service}

\affiliate{icu}{International Christian University}
\affiliate{ist}{Institute of Science Tokyo}
\affiliate{intec}{INTEC Inc.}
\affiliate{ntteast}{NTT EAST}
\affiliate{tsukuba}{University of Tsukuba}
\affiliate{ipa}{Information-Technology Promotion Agency, JAPAN }
\paffiliate{meta}{Meta Platforms, Inc.}

\author{Keisuke Ishibashi}{icu}[ikeisuke@icu.ac.jp]
\author{Xuliang Deng}{icu,meta}
\author{Yoshiaki Kitaguchi}{ist}
\author{Kenichi Nagami}{intec}
\author{Ichiro Mizukoshi}{ntteast}
\author{Akira Sato}{tsukuba}
\author{Daiyu Nobori}{ipa}

\begin{abstract}
With the increasing need for remote work, especially since the COVID-19 era, Remote Desktop Services (RDS) have become widely used. Because interactive RDS usage depends heavily on communication quality, 
some studies have investigated the relationship between QoS metrics and user activity in RDS.
However, these works have been conducted in experimental environments, where the number of samples is limited and may not reflect real-world usage.  Consequently, the relationship between temporal fluctuations in QoS and user activity remains underexplored. 

This paper investigates the relationship between QoS statistics and user activity using real-world usage logs of an RDS, Thin-Telework System. We analyze time-series data of round-trip time (RTT), the number of sent packets, and the number of received bytes per user. Notably, we find that not only the average RTT but also its temporal fluctuation (e.g., standard deviation over time) and its instantaneous deviation from the mean are significantly associated with user activity.
From the users' perspective, these features correspond to QoS unpredictability and the prediction error, respectively, and may provide insights into psychological mechanisms underlying user behavior.
\end{abstract}


\maketitle

\section{Introduction}

Remote work using virtual desktop solutions, such as Remote Desktop Services (RDS),  Desktop as a Service (DaaS) and Virtual Desktop Infrastructure (VDI), has become widespread, especially since the COVID-19 era, because these technologies enable off-site work with relatively low security risk (Hereafter, we use the term RDS for all the above services).
When employees run software on their own or company-issued PCs off-site, sensitive data may leak through cached files or local logs. In contrast, RDS transmits only graphical display frames, so the main remaining risk is eavesdropping, which can be mitigated with encrypted channels. Thanks to these advantages, the global RDS market grew to USD 27.62 billion in 2023~\cite{RDSmarket}.

Despite these benefits, some users suffer degraded interactive performance and user experience, when communication quality, or Quality of Service (QoS), is low. Smooth RDS operation depends on QoS because a wide variety of tasks, document editing, slide creation, chatting, and online meetings, are streamed through the network.
Tolia et al. investigated how network performance limits thin-client usability and found that latency poses a greater challenge than bandwidth; users start noticing lag when round-trip time (RTT) exceeds $150$~ms \cite{tolia2006}. Taylor et al. empirically evaluated the impact of network latency and different remote-desktop delivery methods on user performance and behavior~\cite{taylor2015}. Burke \& Figueroa investigated the relationship between QoS and QoE for RDS and reported that latency sensitivity is application-dependent; for example, text editing satisfaction declined sharply once latency rose above $60$~ms~\cite{burke2021}.

While previous studies have clarified users' QoE under specific QoS conditions in detail, they were all conducted in controlled experimental environments. In recent years considerable progress has been made in assessing the impact of QoS on QoE, user activity, and user engagement in real-world settings, for example in large-scale video streaming services~\cite{Moldovan2019}, \cite{Dobrian2011}, \cite{Krishnan2012imc}. By analyzing real-world usage data, these studies have revealed genuine reductions in user activity during periods of QoS degradation, effects that controlled lab experiments often fail to capture. However, owing to the distributed nature of RDS across clients and servers, measuring a large number of client–server pairs via network measurement has remained difficult.

\begin{figure}[tb]
\centering
\includegraphics[width=0.85\hsize]{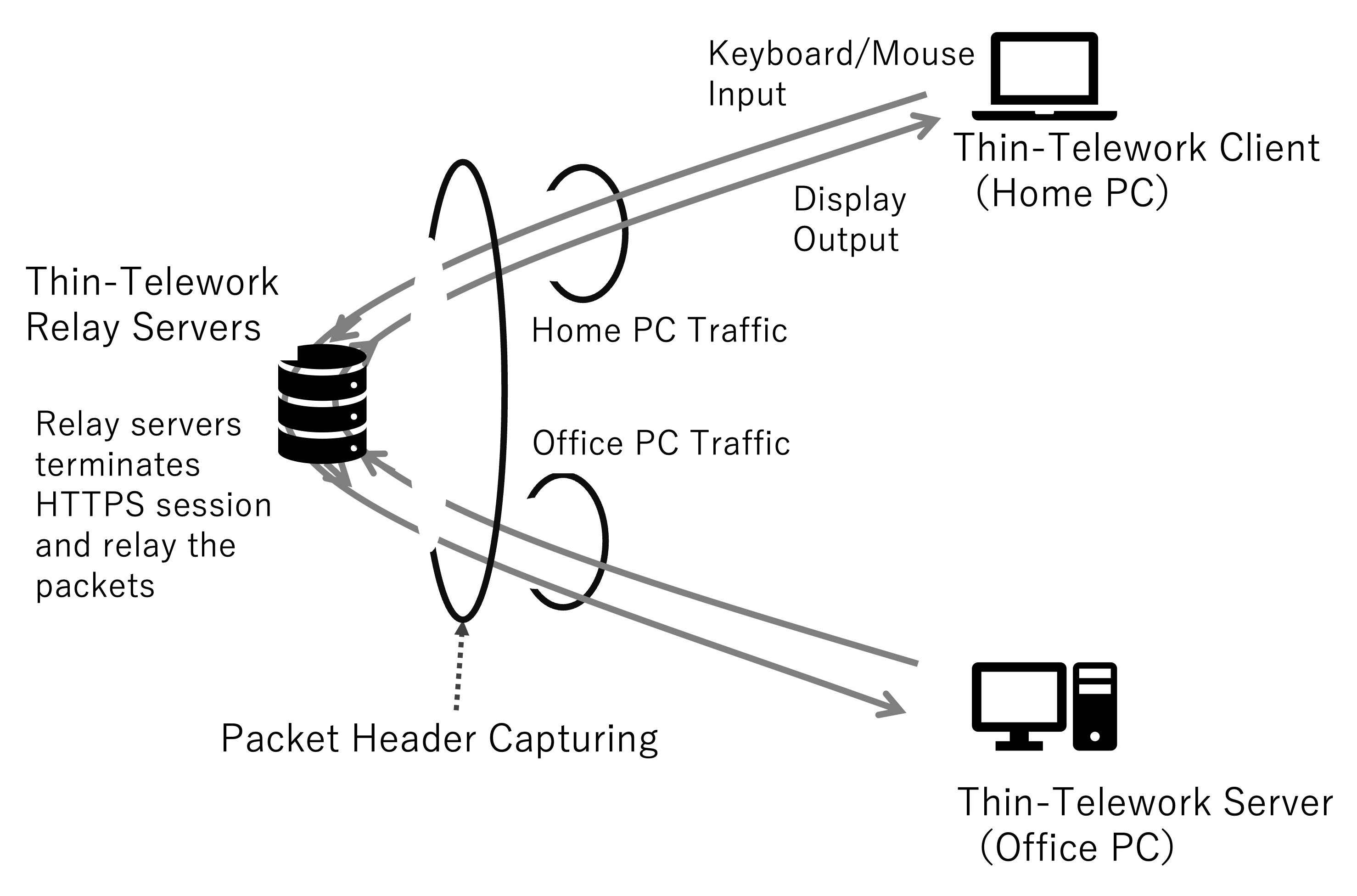}
\caption{Measurement Environment}
\label{fig:CapturingPoint}
\end{figure}

In this paper, we investigate the relationship between QoS and user communication behavior in the Thin-Telework System, an RDS relaying platform developed in 2020 to facilitate remote work during the COVID-19 era~\cite{thintelework}. The service relays RDS traffic as depicted in Figure~\ref{fig:CapturingPoint} and reached $671,929$ users by March 2026. We extracted time-series data of latency (RTT), the numbers of sent packets,  and the number of received bytes from anonymized access logs. 
We then examine how RTT relates to two network-level proxies for user activity: the number of sent packets, which primarily reflects user-generated input such as keystrokes and mouse movements, and the number of received bytes, which consists mainly of graphical frames returned from the office PC. While these metrics do not directly measure usability or subjective QoE, they provide large-scale, objective evidence of how QoS conditions correlate with observable user behavior—a perspective that prior laboratory studies cannot offer.

 In addition to examining the relationship between average RTT and user activity, we also investigate the relationship between other RTT statistics and user activity, 
 and find that both temporal fluctuation and instantaneous deviation from the mean are associated with reduced usage.
Specifically, we calculate three metrics, Exponential Moving Average (EMA), Exponential Moving Standard Deviation (EMSD), and the difference from the previous EMA value (Diff).

Our main contributions are as follows:
\begin{itemize}
\item We provide a large-scale observational analysis of QoS statistics and network-observable user activity in a real-world RDS, using logs from 19,819 users and 39,737,619 active one-minute slots.
\item Beyond average RTT, we examine history-aware QoS statistics, namely EMA, EMSD, and Diff, and show that temporal variability and prediction-error-like features are more strongly associated with user activity when the mean RTT is below 100 ms.
\item We quantify the relative importance of these features using LightGBM and SHAP values, and show that EMSD and Diff are stronger predictors of user activity than EMA. 
\end{itemize}

Rather than measuring subjective QoE directly, this work complements prior laboratory studies by providing large-scale observational evidence on network-observable user activity.

The rest of the paper is organized as follows.
Section~\ref{sec:related} reviews related work on QoE evaluation for RDS, effect of QoS on QoE in the real world, and QoS predictability.  Section~\ref{sec:measurement}  describes our measurement environment and dataset. Section~\ref{sec:analysis} presents our analysis of QoS statistics, user activity, and their relationships. Section~\ref{sec:limitations} discusses the results and their limitations. Finally, Section~\ref{sec:conclusion} summarizes the findings and outlines future work.

\section{Related Works}
\label{sec:related}

\subsection{Effect of QoS on User Activity in the Wild}
Considerable progress has recently been made in studying the effect of QoS on QoE, user activity, and user engagement ``in the wild,'' that is, how engagement, or users' service usage, decreases when QoS degrades~\cite{Moldovan2019}\cite{Dobrian2011}\cite{Krishnan2012imc}\cite{Raca2018}\cite{Plakia2020}\cite{Koto}\cite{Poggi2011}.
One representative behavioral change is abandoning an application such as video streaming: users tend to stop watching when playback stalls (i.e., buffering), which is often induced by increased packet loss or network latency. For example, Dobrian et al. reported that a 1\% increase in buffering ratio can reduce user engagement by more than three minutes during a 90-minute live event~\cite{Dobrian2011}. 

Beyond video streaming, numerous studies have examined web transactions. For instance, a $500$~ms latency increase results in a $20$\% drop in traffic~\cite{linden2006}, and a $100$~ms increase can cause a 7\% decrease in conversion rates~\cite{morton2017}.
Thakkar et al. used a quasi-experimental approach for an online e-mail service and found that raising latency to 500 ms reduced user activity by 12\%.

Unfortunately, such large-scale field studies have not been conducted for RDS. Moreover, most existing works focus on average latency and neglect finer-grained statistics such as temporal deviation or trend.

\subsection{QoE and User Activity in RDS}
In contrast, prior RDS studies have mainly relied on controlled laboratory experiments.
These studies clarify latency sensitivity and subjective user perception, but they do not provide large-scale observational evidence from real-world deployments.

Tolia et al. measured how network performance limits thin-client usability and showed that latency poses a greater challenge than bandwidth; users begin to notice lag when RTT exceeds approximately $150$~ms~\cite{tolia2006}. Taylor et al. empirically evaluated the impact of network latency and different remote-desktop delivery methods on user performance and behavior~\cite{taylor2015}. In their experiment, thin-client users performed three tasks, document editing,  photo editing  and a non-interactive game, under latencies of $30$, $60$, $120$, and $240$~ms. User satisfaction for document editing dropped sharply, whereas satisfaction for photo editing and game remained moderate even at $240$~ms.
Burke and Figueroa further reported that latency sensitivity is application-dependent, with text editing satisfaction declining sharply once latency exceeds $60$ ms~\cite{burke2021}.

% Burke \& Figueroa investigated the relationship between QoS and QoE for RDS and reported that latencies above 60 ms were rated very poorly by users~\cite{burke2021}.

Arellano-Uson et al. developed a method to estimate application-level interactive latency, which is directly correlated with users' QoE~\cite{uson2021}, and evaluated the relationship between this latency and the network-level RTT~\cite{uson2023} for 47 users. They concluded that, when RTT is small, it is not a good indicator of application-level latency because of the processing delay in local computers, but that once RTT exceeds 15 ms, it becomes the primary component of interactivity time.

Although these studies provide detailed insights into QoE under specific QoS conditions, they were conducted in controlled laboratory environments. Consequently, the findings may not fully capture real-world usage and rely largely on questionnaire-based subjective assessments. This work complements prior QoE studies by providing large-scale observational evidence of how QoS statistics relate to user activity in real-world RDS usage.

%\subsection{Effect of QoS on QoE, User Activity, and User Engagement}

\subsection{QoS Predictability}
Temporal trends or deviations in QoS directly influence its predictability, which in turn strongly affect user engagement and QoE. 

For video streaming, low jitter is essential to prevent receiver-buffer overflow or underflow, both of which cause playback stalls. In adaptive streaming, the bit-rate and buffer size are continually adjusted to QoS (e.g. latency or throughput)~\cite{Stockhammer2011acm}\cite{kimura2021}\cite{seufert2015}; consequently,  prediction errors can severely degrade QoE. 

Beyond application-level control, fluctuating QoS reduces predictability from the human user's perspective and thus has psychological consequences. Sabet et al. showed that online-game players can adapt to a certain amount of delay, as long as that delay remains stable~\cite{Sabet2022}.
Conversely, when QoS is unpredictable, QoE declines; the same study notes that “regardless of performance, frequent delay switching annoys gamers.” In a different domain,
Obafemi et al. argue that QoE models for Voice over Internet Protocol (VoIP) should incorporate jitter to one-way delay~\cite{Obafemi2011}.

While prior work has suggested that QoS instability affects user experience across various services, a systematic, data-driven comparison of how mean latency, its variability, and instantaneous prediction error jointly predict user behavior, at scale and in real deployments, remains lacking. 

Motivated by these findings, we investigate how detailed QoS statistics are associated with user activity in the wild.
\section{Measurement Environment}
\label{sec:measurement}

\subsection{Thin-Telework System}

We utilize usage data from an actual RDS platform, the Thin-Telework System. Thin-Telework System was developed by Information-technology Promotion Agency (IPA), Japan  and NTT EAST, and entered service in April, 2020 to provide easy, secure remote access during the COVID-19 era~\cite{thintelework}. 

Users install server software to their office computers and register them to the Thin-Telework System in advance. When they begin remote work, they connect from their home computer to the Thin-Telework relay server via a client application or web browser. The relay server authenticates each user and forwards packets between the home and office PCs. All traffic is carried over an SSL-VPN tunnel encrypted with TLS 1.3, so users do not need to modify the office firewall to accept inbound connections while security is maintained.

% As of March 2026, the system has $671,929$ registered users, of whom $75,612$ were active within the past 30 days.

\subsection{Measurement}

We deployed a packet-capture server in front of the Thin-Telework relay servers and recorded the headers of packets exchanged between the relay and office PCs, and the relay and the home computer (Fig.~\ref{fig:CapturingPoint}). 
Capture was performed with administrative permission. Packet payloads were not stored; IP addresses were anonymized and used to associate packets with individual TCP connections and to infer the user's country of origin. We analyze one week of data collected from 11–17 December 2023.

RTT was calculated between the capture server and each client PC (home PCs and office PCs). We matched the sequence numbers of packets sent from the relay servers with the acknowledgment numbers of the corresponding packets returned by the client PCs, and defined RTT as the timestamp difference between each matched pair. To avoid overestimation due to the Delayed ACK mechanism~\cite{rfc_delayedack}, packet pairs separated by more than $25$~ms were discarded.

\begin{figure}[tb]
\centering
\includegraphics[width=0.8\hsize]{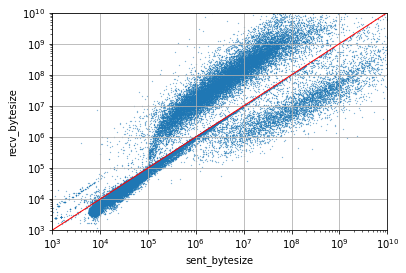}
\caption{Scatter plot of total sending bytes and receiving bytes for each client PCs. The red line shows the points where sending bytes and receiving bytes are equal. We can recognize three clusters; sending bytes larger than $10^6$ and under the red line, receiving bytes larger than $10^6$ and above the red line, sending bytes smaller than $10^6$ and near the red line.}
\label{fig:ByteSentVsByteReceived}
\vspace{0mm}
\end{figure}

Because packets from both home and office PCs traverse the capture point, we first separate the traffic into three groups: (1) office PCs (large upstream byte counts), (2) home PCs (large downstream byte counts), and (3) inactive clients ($\leq 10^6$ bytes in both directions), as illustrated by the scatter plot of sent versus received bytes in Fig.~\ref{fig:ByteSentVsByteReceived}. 
The threshold of $10^6$ bytes was chosen to exclude TCP connection setup and keep-alive traffic that does not reflect sustained RDS sessions; varying this threshold by an order of magnitude in either direction did not materially change the composition of the identified clusters.
In RDS, home PCs mainly transmit user input events, whereas office PCs return screen images; therefore, home-PC flows exhibit far fewer sent than received bytes.
%Our subsequent analysis focuses on home-PC clients. As proxies for user activity, we use the number of packets sent (user input) and the number of bytes received (application response). 
To exclude automated or overseas usage, we restrict the dataset to home PCs located in Japan, identified via a Geo-IP service~\cite{geoIP}. The resulting dataset contains $19,819$ unique home-PC IP addresses.

There exist many QoS metrics other than RTT, the reason why we choose RTT is explained in the next subsection, as well as the rationale for using the number of packets sent and bytes received as proxies of user activity.

We then aggregate the packet data into $60$-second slots and compute the mean RTT within each slot. This yields an RTT time series of length $10,080 (= 60 \times 24 \times 7)$ for every user, which we use as the primary QoS indicator. For user-activity indicators, we use the number of packets sent and the number of bytes received by each home PC during the same $60$-second slot.
We then extract time slots that contain at least one RTT sample and a non-zero packet count.

With one week of data, there are $10,080$ one-minute slots for each of the $19,819$ users, and by excluding inactive slots we obtain $39,737,619$ active time-slot samples. This sample size is sufficient for our analysis of short time-scale phenomena at the one-minute level. On the other hand, longer time-scale effects, such as seasonal effects or long-term trends in which user context may change, are left for future work.

\subsection{Metrics of User Activity and QoS}
\label{UserQoSMetrics}

\subsubsection{User Activity Metrics}

\begin{figure}[tb]
\centering
\includegraphics[width=0.85\hsize]{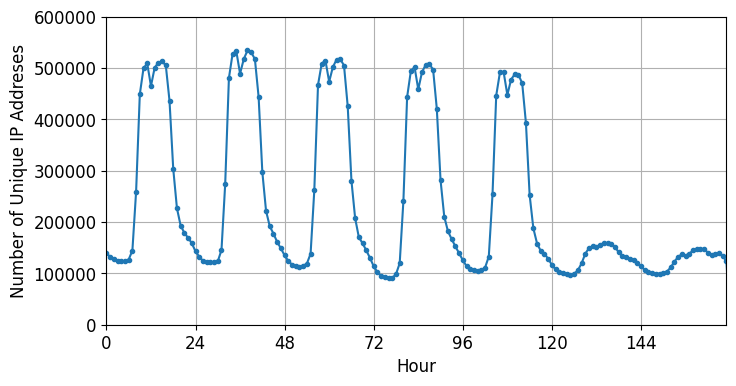}
\caption{Time Series of Number of Unique IP Addresses. X-axis starts with the beginning of the measurement, period, Monday 0 a.m. December 11th, 2023.}
\label{fig:active_user_time_series}
\end{figure}

In the target RDS use case (the Thin-Telework System), the main purpose of use is expected to be office work. Figure~\ref{fig:active_user_time_series} shows the number of unique client IP addresses that appear in each hour during the one-week measurement period. It exhibits a typical daily pattern, with a clear increase between 9 a.m. and 5 p.m. from Monday to Friday, which indicates that the system is mainly used for office work. 

In the office applications used on this system, users provide input
via keyboard and mouse and then wait for the graphical frames returned from the office PC. Although these input and output events are not directly observed in the network, the number of packets sent and the number of bytes received can be regarded as proxies for these events. In Appendix~\ref{A}, we conduct a small experiment to confirm that  the number of packets sent and the number of bytes received depends on the user activity for office applications.

\subsubsection{QoS Metrics}
\label{qosmetrics}
Various QoS metrics may be relevant to remote interactive applications. In this study, we focus on RTT as the primary QoS metric for analyzing user activity for the following reasons:
\begin{itemize}
\item Previous laboratory studies have reported that latency is the primary QoS factor for such remote desktop applications~\cite{tolia2006},~\cite{taylor2015}, and~\cite{burke2021}.
\item RTT reflects both delay and jitter, because jitter is the temporal variation of delay and can be captured through temporal fluctuations of RTT. In our analysis, EMSD is introduced to quantify this variability.
\item Since the RDS sessions run over TCP, packet loss is recovered by retransmissions and mainly appears as additional delay due to the retransmission or as reduced throughput due to congestion control. In other words, the impact of moderate loss is largely reflected in the effective RTT and its variability.
\item The office applications considered here do not appear to be strongly throughput-sensitive, in contrast to video-streaming applications. In addition, the 99th percentile of the number of bytes received in 60 seconds is about 14 MB (see Fig.~\ref{fig:Byte_CCDF}), which corresponds to approximately 1.8 Mbps. Because this value is significantly smaller than typical access-link capacities today, throughput limitations are unlikely to be the primary cause of user activity degradation in our environment. However, a joint analysis of throughput and RTT for throughput-sensitive applications is an important topic for future work.
\end{itemize}

\section{Analysis Results}
\label{sec:analysis}

\subsection{Basic Statistics}

% \begin{figure*}[htb]
% %    \centering
%   \begin{minipage}{0.5\hsize}
%      \centering
% \includegraphics[width=0.8\columnwidth, height=0.36\columnwidth]{figs/Packet_CCDF.png}
% \caption{CCDF of the number of sent packets from home PC in \\60 seconds}
% \label{fig:Packet_CCDF}
%   \end{minipage}
%   \begin{minipage}{0.5\hsize}
%    \centering
% \includegraphics[width=0.8\columnwidth, height=0.36\columnwidth]{figs/Byte_CCDF.png}
% \caption{CCDF of the number of received bytes by home PC in \\60 seconds}
% \label{fig:Byte_CCDF}
%   \end{minipage}
% \end{figure*}

\begin{figure}[tb]
     \centering
\includegraphics[width=0.8\columnwidth, height=0.36\columnwidth]{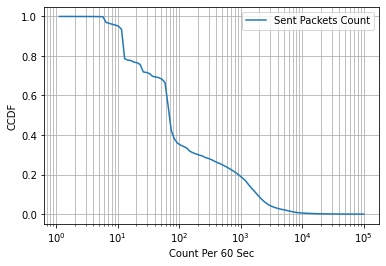}
 \caption{CCDF of the number of sent packets from home PC in \\60 seconds}
 \label{fig:Packet_CCDF}
 \end{figure}

\begin{figure}[tb]
     \centering
\includegraphics[width=0.8\columnwidth, height=0.36\columnwidth]{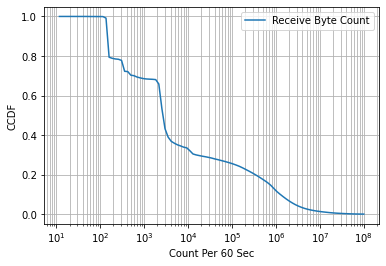}
 \caption{CCDF of the number of received bytes by home PC in \\60 seconds}
 \label{fig:Byte_CCDF}
   \end{figure}

First, we present basic traffic and QoS statistics for home PCs. Figures~\ref{fig:Packet_CCDF} and~\ref{fig:Byte_CCDF} show the complementary cumulative distribution functions (CCDFs) for the number of packets sent and the number of bytes received per 60-second slot for all home-PC clients. Although the two CCDFs exhibit a similar shape, their $x$-ranges differ.

Both CCDFs reveal step-like drops. For example, the CCDF of sent packets falls sharply around 12 and 60 packets, reflecting periodic keep-alive messages sent every five seconds or one second, respectively. 
Despite focusing on active home PCs, a considerable number of time slots contain no user operations or think time. 
Excluding these keep-alive periods, the number of sent packets spans $10^2$  to $10^4$, whereas received-bytes counts range from $10^3$ to $10^7$.
Among active users, the mean and median counts of sent packets are $1,729$ and $1,072$, respectively; for received bytes they are $2,149,247$ and $468,423$.

\begin{figure}[tb]
\centering
\includegraphics[width=0.8\columnwidth, height=0.36\columnwidth]{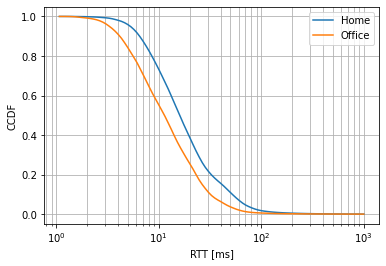}
\caption{CCDF of average RTT between relay servers and both Home and Office PC in 60 seconds}
\label{fig:RTT_CCDF}
\vspace{0mm}
\end{figure}

Figure ~\ref{fig:RTT_CCDF} depicts the CCDFs of home-PC RTT. RTT values range from $2$~ms to $300$~ms, and $70$\% of time slots fall between $10$ and $100$~ms. We also plot office-PC RTT, which are markedly smaller.
This is because most office PCs use wired connections and are geographically close to the Tokyo relay servers, whereas home PCs are distributed nationwide and may use wireless access. The average RTTs are $25$~ms for home PCs and $8$~ms for office PCs.

Although users perceive the sum of relay-to-home and relay-to-office RTTs, the latter is relatively small and stable, we focus on the home-PC RTT as the primary QoS metric in subsequent analyses.

\subsection{Relationship between RTT Statistics and User Activity}

\subsubsection{Association between Instantaneous RTT and User Activity}

\begin{figure}[tb]
     \centering
\includegraphics[width=0.8\columnwidth, height=0.36\columnwidth]{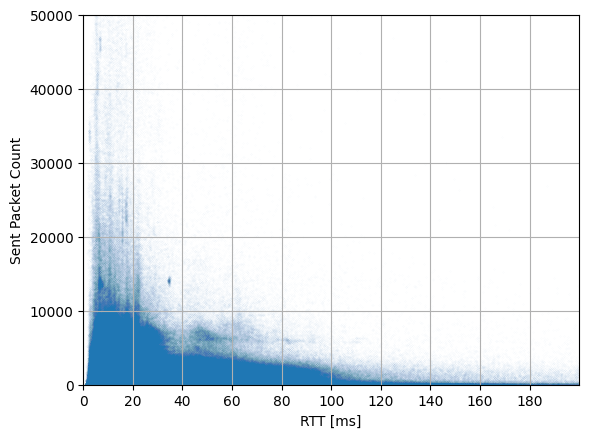}
\caption{Scatter plot for RTT and the number of sent packets}
\label{fig:ScatterSentPacketsVsRTT}
  \end{figure}
\begin{figure}[tb]
   \centering
\includegraphics[width=0.8\columnwidth, height=0.36\columnwidth]{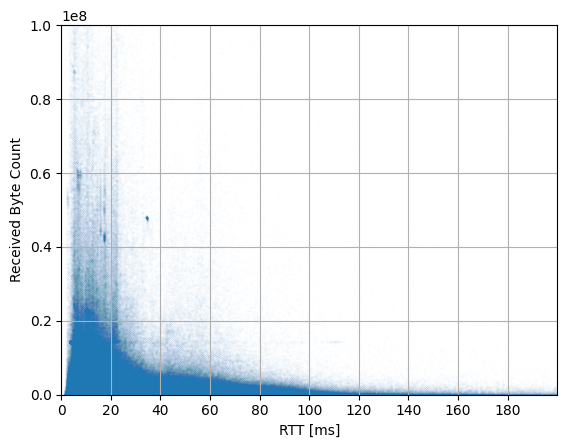}
\caption{Scatter plot for RTT and the number of received bytes}
\label{fig:ScatterByteReceivedVsRTT}
\end{figure}

\begin{figure}[tb]

     \centering
\includegraphics[width=0.8\columnwidth, height=0.36\columnwidth]{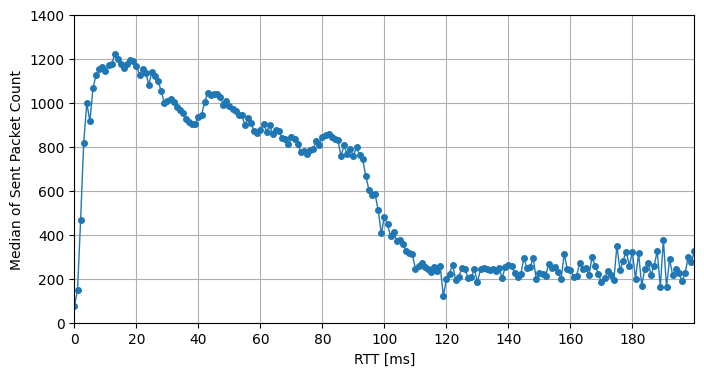}
\caption{Median of number of sent packets for binning RTT}
\label{fig:SentPacketsVsRTT}
  \end{figure}
\begin{figure}[tb]
   \centering
\includegraphics[width=0.8\columnwidth, height=0.36\columnwidth]{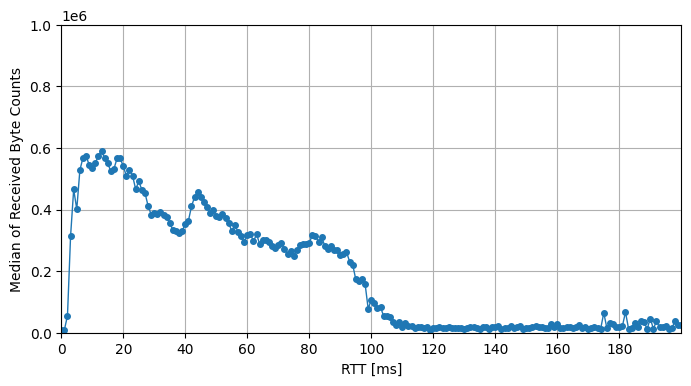}
\caption{Median of number of received bytes for binning RTT}
\label{fig:ByteReceivedVsRTT}

\end{figure}

We next investigate how RTT statistics are associated with user activity.
Figures~\ref{fig:ScatterSentPacketsVsRTT} and \ref{fig:ScatterByteReceivedVsRTT} plot scatter diagrams of RTT versus the number of packets sent, bytes received, respectively; each point represents one 60-second time slot for a client.  As RTT exceeds $30$~ms, outliers with exceptionally high packet or byte counts disappear, and once RTT surpasses $100$~ms, the majority of points with moderate activity also vanish.

However, because the number of observed slots also decreases as average RTT increases, it is unclear whether this decline stems from reduced user activity or merely from fewer samples.
To disentangle these factors, we group the data into $1$~ms RTT bins and compute the median activity for each bin\footnote{We use the median to mitigate outlier effects, the mean would be influenced by rare bursts.}.
Figures~\ref{fig:SentPacketsVsRTT} and  ~\ref{fig:ByteReceivedVsRTT} show the median numbers of packets sent and bytes received per bin. The previous trend, sharp drops beyond $30$~ms and $100$~ms, remain evident at the median level, confirming that they are not results of sample-count reduction. 
These findings align with earlier studies~\cite{tolia2006},~\cite{taylor2015}.
There is a sharp drop below $10$~ms, which we attribute to automated monitoring or keep-alive connections from within the same AS rather than genuine interactive user sessions; these slots are retained in the analysis because excluding them does not materially affect the trends observed above $10$~ms.

%\end{multicols}

\subsubsection{Association between RTT History and User Activity}

\begin{figure}[tb]
     \centering
\includegraphics[width=0.8\columnwidth, height=0.36\columnwidth]{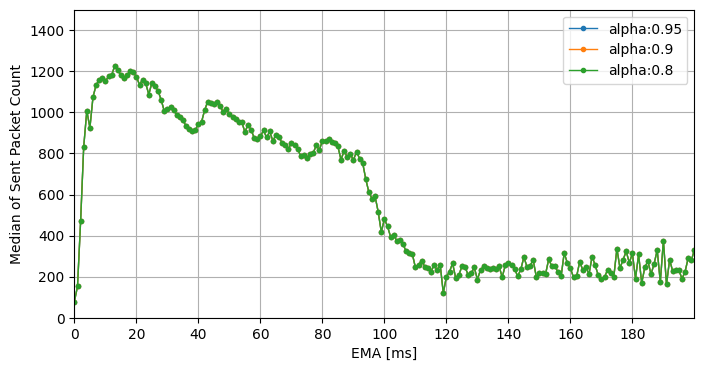}
\caption{Median of number of sent packets for binning \\ Exponential Moving Average (EMA)}
\label{fig:SentPacketsVsEWM}
  \end{figure}
\begin{figure}[tb]
   \centering
\includegraphics[width=0.8\columnwidth, height=0.36\columnwidth]{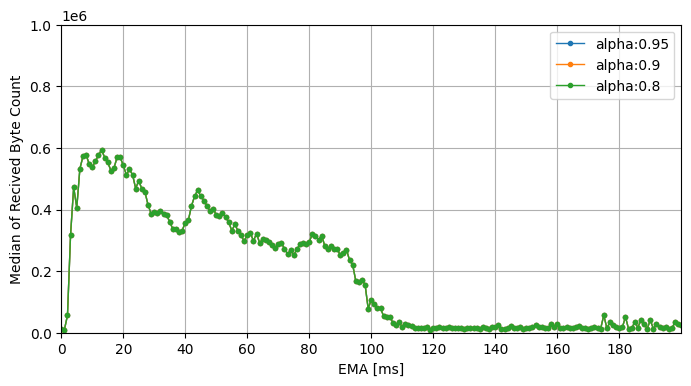}
\caption{Median of number of received bytes for binning \\ Exponential Moving Average (EMA)}
\label{fig:ByteReceivedVsEWM}
\end{figure}

\begin{figure}[tb]
     \centering
\includegraphics[width=0.8\columnwidth, height=0.36\columnwidth]{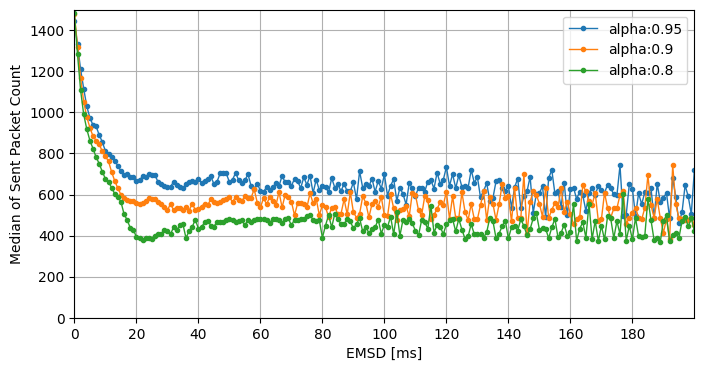}
\caption{Median of number of sent packets for binning\\ Exponential Moving Standard Deviation (EMSD) }
\label{fig:SentPacketsVsEWSD}
  \end{figure}
\begin{figure}[tb]
   \centering
\includegraphics[width=0.8\columnwidth, height=0.36\columnwidth]{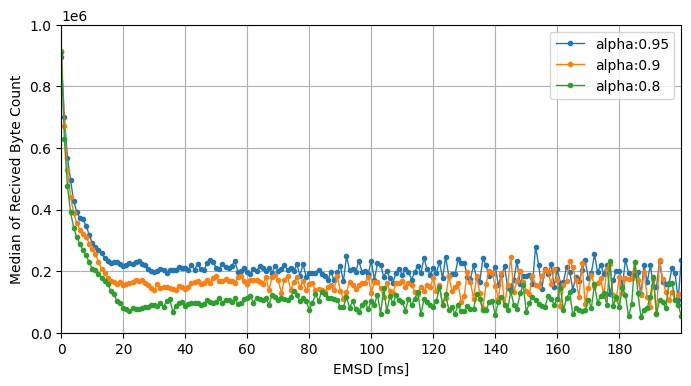}
\caption{Median of number of received bytes for binning \\ Exponential Moving Standard Deviation (EMSD)}
\label{fig:ByteReceivedVsEWSD}
\end{figure}

\begin{figure}[tb]
     \centering
\includegraphics[width=0.8\columnwidth, height=0.36\columnwidth]{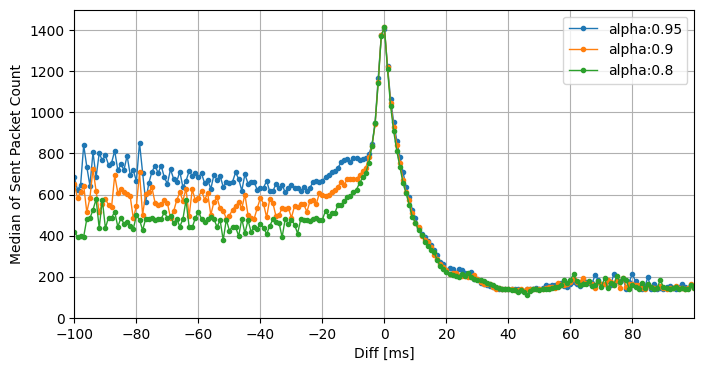}
\caption{Median of number of sent packets for binning \\ Difference with Exponential Moving Average (Diff)}
\label{fig:SentPacketsVsDiff}
  \end{figure}
\begin{figure}[tb]
   \centering
\includegraphics[width=0.8\columnwidth, height=0.36\columnwidth]{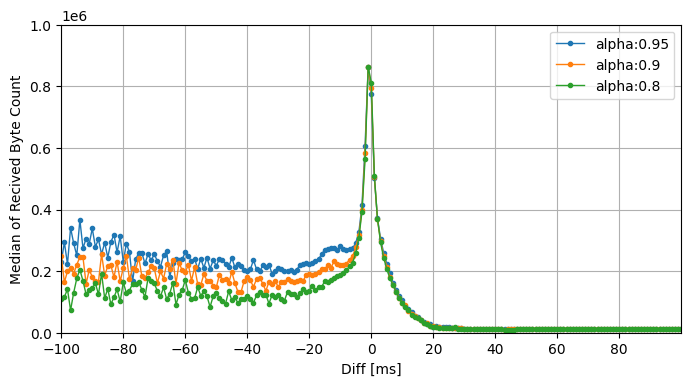}
\caption{Median of number of received bytes for binning \\ Difference with Exponential Moving Average (Diff)}
\label{fig:ByteReceivedVsDiff}
\end{figure}

Thus far, we have examined the instantaneous effect of RTT on user activity by analyzing data within the same $60$ second slot. 
However, user behavior may also depend on preceding QoS conditions.
To capture such history effects, we compute three exponentially weighted statistics for each slot: the exponential moving average (EMA, $E_t$), the exponential moving standard deviation (EMSD, $S_t$),  and the one-step prediction error (Diff, $D_t$).
They are defined recursively as
\begin{align}
E_t &= \alpha R_{t} + (1 - \alpha ) E_{t-1} &\\
S_t &:= \sqrt{V_t}\nonumber\\ 
 V_t &:= \alpha (R_{t} - E_t)^2 + (1 - \alpha ) V_{t-1} \\
D_t &= R_t - E_{t-1},
\end{align}

where $R_t$ is the mean RTT in slot $t$. To capture short-term QoS history and to examine the effect of the smoothing factor, we use $\alpha = 0.8, 0.9, 0.95$, and confirm that our main observations are robust across these values.
In the literature, EMA is used to model human expectation for continuous stimuli~\cite{EMA}; accordingly, we interpret EMA as the user's RTT expectation or prediction, EMSD as RTT instability or unpredictability, and Diff as the instantaneous prediction error.

Figures~\ref{fig:SentPacketsVsEWM},~\ref{fig:ByteReceivedVsEWM},~\ref{fig:SentPacketsVsEWSD},~\ref{fig:ByteReceivedVsEWSD},~\ref{fig:SentPacketsVsDiff}, and~\ref{fig:ByteReceivedVsDiff} plot the medians of packets sent and bytes received against binned EMA, EMSD, and Diff (bin width = $1$~ms). 
The EMA curves resemble those for the raw RTT mean, whereas both EMSD and Diff exhibit distinct patterns: (i) as EMSD increases up to around $20$~ms, activity drops rapidly and then plateaus; (ii) as Diff increases beyond approximately $20$~ms, activity again drops rapidly. Interestingly, values below about $-2$~ms are also associated with reduced activity. One possible explanation is that users prefer stable latency and do not immediately take advantage of sudden improvements; alternatively, large negative Diff may correspond to transitions from previously poor conditions, where user activity has already been discouraged. In contrast, a Diff of about $-1$~ms indicates a small improvement with tolerable latency variability, and user activity remains high. The above observations are common for all $\alpha$s.
Hereafter, we select $\alpha = 0.9$ as the representative value because it corresponds to an effective memory window of approximately $10$ slots ($10$ minutes), which aligns with plausible timescales of user adaptation to network conditions in interactive desktop tasks.

\begin{figure*}[t] 
    \includegraphics[keepaspectratio, scale=0.365]{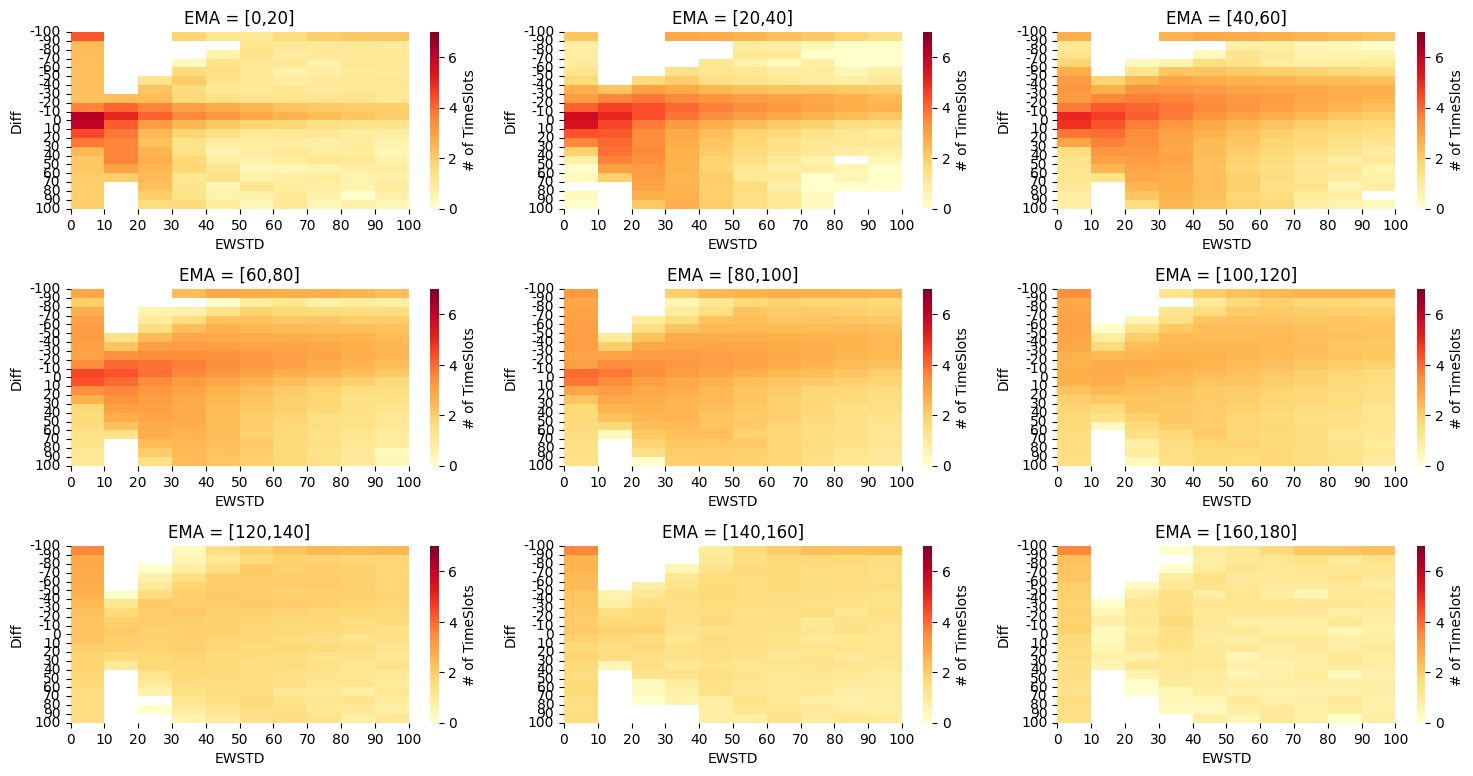}
    \centering
    \caption{Heat-maps for the number of time slot counts for each EMSD and Diff bins, conditioning by EMA. Colors are drawn in log scale.}
  \label{Heatmap_EMA_EMSD_Diff_Count}
      \vspace{0em}
\end{figure*}

%\begin{multicols}{2}
\begin{figure*}[t] 
    \includegraphics[keepaspectratio, scale=0.365]{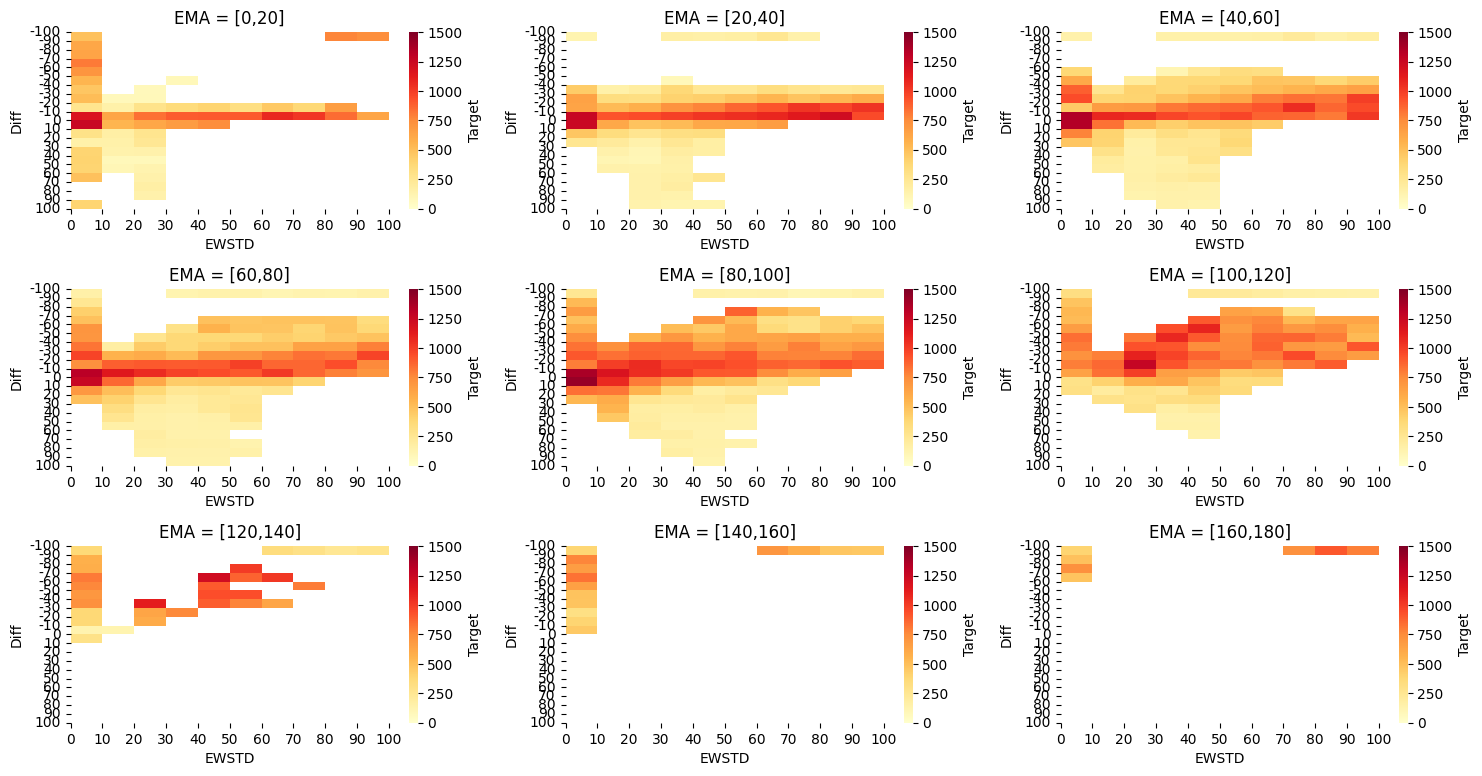}
    \centering
    \caption{Heat-maps for the number of sent packets for each EMSD and Diff bins, conditioning by EMA}
  \label{Heatmap_EMA_EMSD_Diff_SentPkt}
      \vspace{0em}
\end{figure*}

\begin{figure*}[t] 
    \includegraphics[keepaspectratio, scale=0.365]{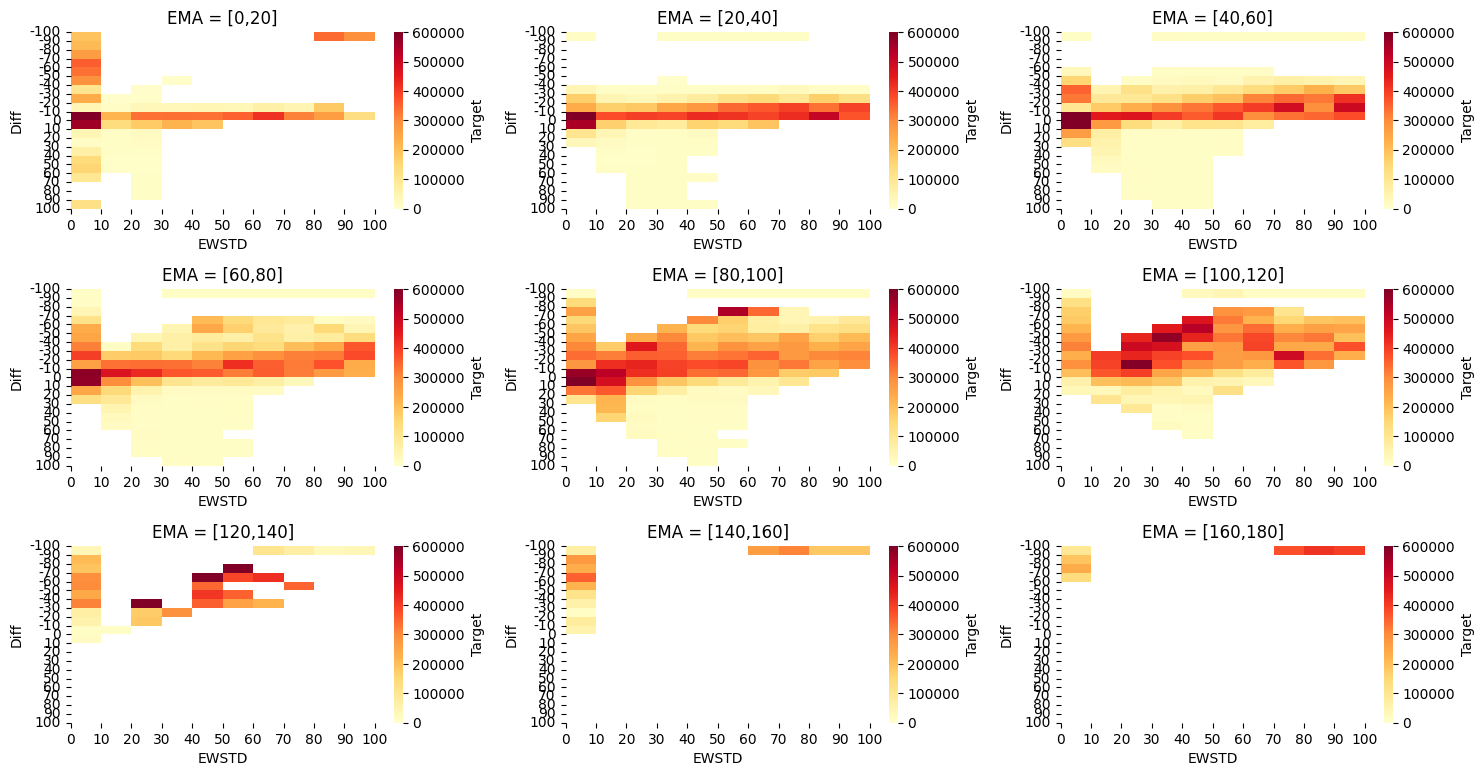}
    \centering
    \caption{Heat-maps for the number of received bytes for each EMSD and Diff bins, conditioning by EMA}
  \label{Heatmap_EMA_EMSD_Diff_RecvByte}
      \vspace{0em}
\end{figure*}

To explore joint effects, we generate heat-maps of EMSD versus Diff for several EMA ranges (we use $20$~ms bin for EMA in these heat-maps).
Figure~\ref{Heatmap_EMA_EMSD_Diff_Count} shows heat-maps of time-slot counts. Note that the colors are drawn on a logarithmic scale. For heat maps with EMA smaller than $100$ ms, there is a vertical stripe at EMSD $< 10$ ms and a horizontal stripe in the middle of the heat map, consistent with the marginal plots.
Additional heat maps for sent packets and received bytes are shown in Figs.~\ref{Heatmap_EMA_EMSD_Diff_SentPkt} and  ~\ref{Heatmap_EMA_EMSD_Diff_RecvByte}, respectively; cells with fewer than 100 samples are omitted to avoid sparse artifacts.
The high-activity region—concentrated at low EMSD and near-zero Diff—is most clearly visible in the EMA $=[0,20]$ and $[20,40]$ panels, where the bright zone collapses sharply as EMSD exceeds approximately $10$~ms or the absolute values of Diff exceed approximately $5$~ms. This confirms that, at low mean latency, temporal stability rather than the mean level itself is the dominant correlate of user activity.
As EMA increases, the horizontal stripes tilt upward, indicating that larger absolute Diff is tolerated when the baseline RTT is high, mirroring the asymmetry observed in Fig.~\ref{fig:SentPacketsVsDiff}. 
Taken together, these results suggest that, once mean latency is below $100$~ms, its variability (EMSD) and prediction error (Diff) show stronger association with user activity, consistent with psychological and reinforcement-learning studies that emphasize prediction error as a primary driver of human response~\cite{schultz1997}.

\begin{table}[tb]
\centering
\caption{SHAP}
\label{tab:SHAP}
\begin{tabular}{lcc}
\toprule
 Feature & RFI  & RFI\\ 
  &  (Sent Packets) &  (Received Bytes)\\
\midrule
 EMA  & $21.0$\%  & $23.5$\%\\ \hline
 EMSD& $31.8$\% & $41.9$\% \\ \hline
 Diff& $47.2$\% & $33.4$\% \\ \hline
\bottomrule
\end{tabular}
\label{tab:MAE}
\end{table}

\subsubsection{Significance of RTT-derived Features}
To quantify the observations in the previous section, we evaluate the relative importance of the three RTT-derived features, EMA, EMSD, and Diff, in predicting user activity.
We train a LightGBM regressor~\cite{lightGBM} to predict the numbers of packets and bytes, and compute 
SHAP (SHapley Additive exPlanations) values~\cite{SHAP} for each feature.
The SHAP value $\phi_i^{(j)}$ of feature $j$ in sample $i$ is defined as

 \begin{align}
&\phi_i^{(j)} := \sum_{S \subset F\setminus \{j\}}\frac{(|N|-|S|-1)!|S|!}{|N|!} \times &\nonumber \\ 
&\left({\mathbb E}[f(X)|X_{S}=x_{S}, X_j = x_j]-{\mathbb E}[f(X)|X_{S}=x_{S} ]\right),&
\label{eq:SHAP}
\end{align}
where $N$ is the full feature set and $f(\cdot)$ is the prediction function (the LightGBM regressor).
Because SHAP values are computed for each sample, we summarize them by the Relative Feature Importance (RFI):
\begin{align}
R_j := \frac{\sum_i |\phi_i^{(j)}|}{\sum_k\sum_i |\phi_i^{(k)}|}
\label{eq:RFI}
\end{align}

Table~\ref{tab:SHAP} lists the results. 
Surprisingly, EMA has the lowest RFI. 
More precisely, Diff shows a higher importance score for sent packets, whereas EMSD shows a higher importance score for received bytes. One possible explanation is that the number of packets sent mainly reflects user input behavior and may be more sensitive to short-term deviations in RTT, which are captured by Diff. In contrast, the number of bytes received reflects a combination of user input and application-side responses (for example, screen updates), which may depend more on longer time-scale variability, as captured by EMSD.

\section{Discussion and Limitations}

\label{sec:limitations}

\subsection{Activity Proxy and Its Limitations}

We use the number of sent packets and the number of received bytes as coarse, network-level proxies for user activity, as user activity cannot be measured through encrypted network data. These metrics reflect a combination of application type and usage patterns rather than direct observations of user actions.

That said, these proxies have several limitations that should be kept in mind when interpreting the results.
First, they do not directly distinguish between user think time and active input periods. Second, they depend on application-specific behaviors such as batching, compression, and rendering mechanisms. As a result, the same level of user activity may produce different traffic patterns across applications.

If application types or contexts could be inferred, a more fine-grained analysis of user activity would be possible. In particular, separating user think time from active interaction periods would provide deeper insights into user behavior.

\subsection{QoS Metrics}
RTT is estimated from seq/ack matching and is therefore only available during bidirectional packet exchanges. In low-activity slots, fewer matched pairs may be available, potentially introducing a selection bias toward slots with higher activity levels. This may cause the observed RTT distributions in low-activity regimes to be less reliable.

\subsection{Observation Period}

The dataset covers only one week in mid-December. Although this is sufficient for analyzing short-term relationships at the one-minute time scale, it does not capture longer-term variations such as seasonal effects, holiday schedules, or period-specific work patterns. Accordingly, the generality of the results beyond the observed week remains to be validated using longer-term measurements.

\subsection{Causality and Confounders}

\begin{figure}[tb]
\centering
\includegraphics[width=0.85\hsize]{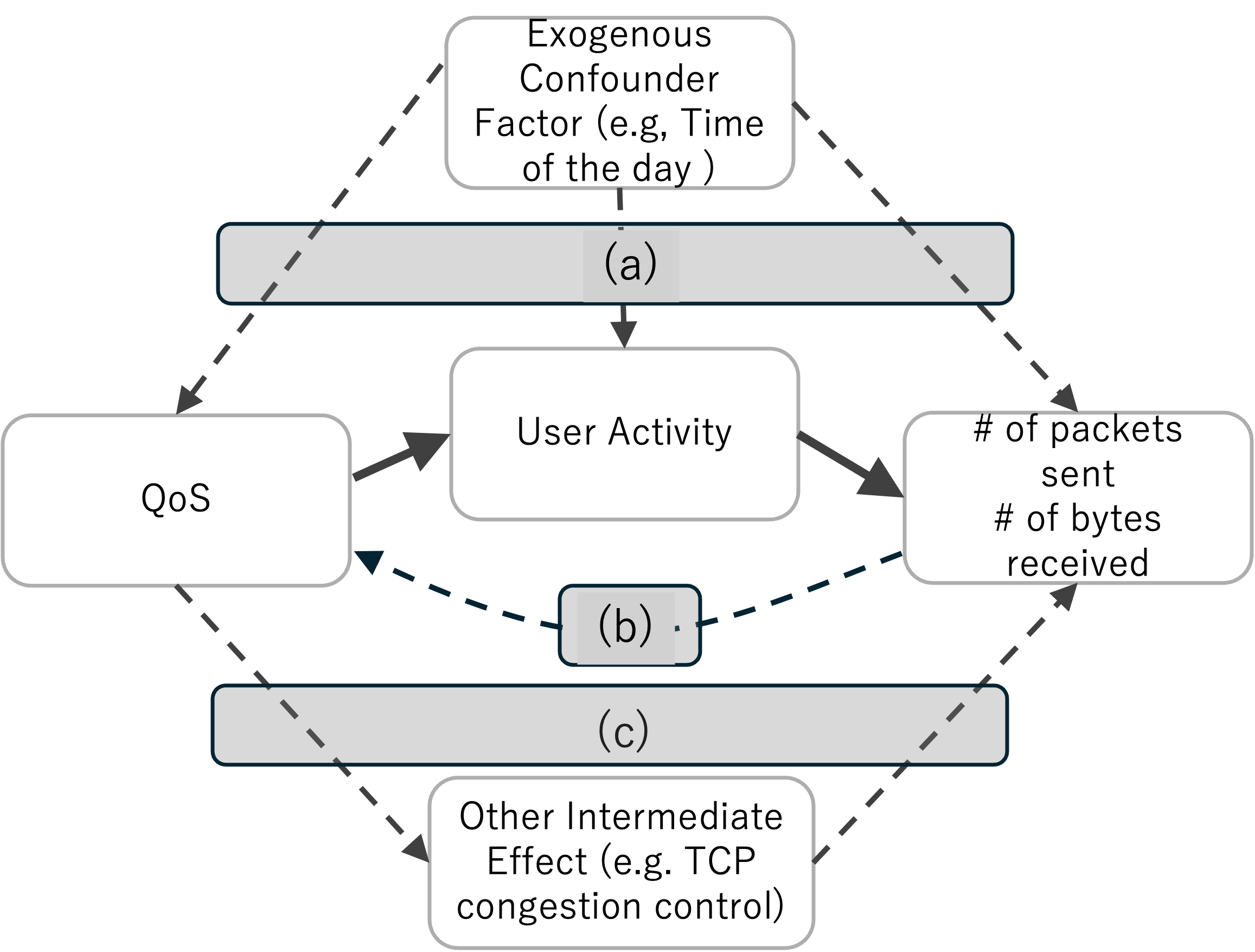}
\caption{Possible causal relations between QoS and packet/byte counts}
\label{fig:CausalDiagram}
\end{figure}

We have confirmed correlations between QoS statistics and packet/byte counts, which we treat as network-observable proxies for user activity. The associations between EMSD, Diff, and user activity are consistent and robust across values of $\alpha$, suggesting that these statistics may serve as useful signals of changes in user activity. Nevertheless, correlation does not imply causation, and several alternative explanations warrant consideration. 
Figure~\ref{fig:CausalDiagram} summarizes the possible relationships, which we discuss in turn.

\begin{enumerate}[label=(\alph*)]
\item \textbf{Confounding factors.}  A common factor may influence both QoS and user activity (and hence the packet/byte counts). Candidates include application type, user context, and time-of-day effects—for example, in the early morning both RTT and user activity tend to be low. If such a factor affects QoS and user activity in the same direction, QoS degradation could spuriously appear to cause reduced user activity. We have not yet identified specific confounders of this kind; investigating them is left for future work.
\item \textbf{Reverse causality.} Rather than QoS degradation reducing user activity, increased packet and byte counts—driven by higher user activity or a change of application—could congest the network and thereby degrade QoS. This would induce a negative correlation from user activity to QoS. Because such reverse causality opposes the positive association we observe, its presence would imply that the true effect from QoS to user activity is even stronger than the observed correlation suggests. However, the average bit rate is low (see below), so network-level congestion of this kind appears unlikely in our environment.
\item \textbf{Intermediate factors.} 
RTT can also affect packet/byte counts through TCP congestion-avoidance algorithms. As discussed in section~\ref{qosmetrics} the average bit rate for an RDS session (about 1.8 Mbps) is significantly smaller than typical available bandwidth today; we therefore expect that TCP congestion avoidance does not constrain the sending rate, even in a high-RTT environment. A joint analysis of throughput and RTT for throughput-sensitive applications nonetheless remains important future work.

\end{enumerate}
In summary, the observed relationships should be interpreted as correlations between QoS statistics and network-observable user activity. A more rigorous causal analysis, using application-level measurements or controlled experiments, is left for future work.

\section{Conclusion}
\label{sec:conclusion}

We analyzed one week of real-world RDS logs and examined how RTT is associated with user activity.
Beyond the mean latency, we introduced two history-aware features, EMSD and Diff, and found that both are strongly correlated with user activity when the mean RTT is below $100$~ms.
In practice, once the average latency is kept within an acceptable range,  monitoring latency stability through EMSD and Diff may offer more informative signals of user experience changes than monitoring mean latency alone.

From the users' perspective, fluctuation represents difficulty in predicting QoS, whereas deviation represents prediction error of QoS. These findings highlight the importance of interpreting QoS statistics through a psychological lens.
We believe that our results can inform more efficient network and application control; if these correlations reflect causal effects, operators should prioritize predictability and minimize fluctuations once the average latency is within an acceptable range.
To our knowledge, this is the first large-scale RDS study that links sub-100 ms temporal QoS statistics to observed user communication behavior.

% Taken together, these findings indicate that future QoE models and network-control schemes should consider prioritizing latency stability and predictability, rather than focusing solely on average latency.

%measure the RDS usage in the wild, and investigate the relationship between QoS (RTT) and user activity. Specifically, we focus not only average but also deviation and temporal difference from average, and found the latter two features significantly affect the user activity. Because deviation and temporal difference can be interpreted as in unpredictability and prediction error, and reward predictability is known to heavily affects on the psychological response~this findings may open the possibility of psychological interpretation QoE analysis. 
%\bibliographystyle{ieicetr}% bib style
%\bibliography{}% your bib database

\vspace{2em}
\begin{thebibliography}{99}% more than 9 --> 99 / less than 10 --> 9


\bibitem{RDSmarket}
Hardware \& Software IT Services, 
``Device as a Service (DaaS) Market Size, Share \& Industry Analysis,''
Fortune Buisiness Inside, 
\url{https://www.fortunebusinessinsights.com/device-as}\\\url{-a-service-market-108000}. [Online, accessed 8-June-2026]


\bibitem{tolia2006}
N.~Tolia, D.~G.~Andersen, and M.~Satyanarayanan, ``Quantifying interactive user experience on thin clients,'' \emph{IEEE Computer}, vol.~39, no.~3, pp.~46--52, Mar.~2006.


\bibitem{taylor2015}
B.~Taylor, Y.~Abe, A.~K.~Dey, and M.~Satyanarayanan, ``Virtual machines for remote computing: Measuring the user experience,'' Carnegie Mellon University Technical Report CMU-CS-15-101, Pittsburgh, PA, Jan.~2015.

\bibitem{burke2021}
A.~Burke and M.~Figueroa, ``Latency perception in cloud-based workspaces and environments,'' \emph{SMPTE Motion Imaging Journal}, vol.~130, no.~7, pp.~31--38, Aug.~2021.

\bibitem{Moldovan2019}
C. Moldovan, F. Wamser, T. Ho\ss feld,
``User Behavior and Engagement of a Mobile Video Streaming User from Crowdsourced Measurements,''  in Proc. 2019 Eleventh International Conference on Quality of Multimedia Experience (QoMEX), June 2019.


\bibitem{Dobrian2011}
F. Dobrian, et al., ``Understanding the Impact of Video Quality on User Engagement,''
\textit{ACM SIGCOMM Computer Communication Review}, Vol. 41, Issue 4, August 2011.


\bibitem{Krishnan2012imc}
S. S. Krishnan and R. K. Sitaraman, ``Video Stream Quality Impacts Viewer Behavior,'' in Proc. ACM Internet Measurement Conference (IMC), November 2012.

\bibitem{thintelework}
NTT EAST--IPA, ``Thin Telework System'',\\ \url{https://telework.cyber.ipa.go.jp/news/}. [Online, accessed 8-June-2026]





\bibitem{Raca2018}
D.~Raca, D.~Leahy, C.~J. Sreenan, and J.~J. Quinlan, ``Incorporating prediction into adaptive streaming algorithms: A QoE perspective,'' in \emph{Proc. ACM Workshop on Network and Operating Systems Support for Digital Audio and Video (NOSSDAV '18)}, Amsterdam, Netherlands, June 2018, pp.~19--24.

\bibitem{Plakia2020}
E.~Plakia, G.~Mylonas, and P.~Papadimitriou, ``Should I stay or should I go: Analysis of the impact of application QoS on user engagement in YouTube,'' \emph{ACM Trans. Multimedia Comput. Commun. Appl.}, vol.~16, no.~3, pp.~1--21, Aug.~2020.

\bibitem{Koto}
H. Koto, N. Fukumoto, S. Niida, H. Yokota, S. Arakawa, and M. Murata, ``Users Reaction to Network Quality During Web Browsing on Smartphones,'' in Proc. 26th International Teletraffic Congress (ITC), September 2014.

\bibitem{Poggi2011}
N. Poggi, D. Carrera, R. Gavalda, and E. Ayguade,
``Non-intrusive Estimation of QoS Degradation Impact on E-Commerce User Satisfaction,''
in Proc. IEEE 10th International Symposium on Network Computing and Applications, August 2011.

\bibitem{linden2006}
G. Linden. Geeking with greg. \url{http://glinden.blogspot.com/2006/11/marissa-mayer-at-web-20.html}, 2021. [Online, accessed 8-June-2026].

\bibitem{morton2017}
R. Morton and T. Barth.
Akamai Online Retail Performance Report: Milliseconds Are Critical,
\url{https://www.ir.akamai.com/news-releases/news-release-details/}\\
\url{akamai-online-retail-performance}\\
\url{-report-milliseconds-are} Apr. 2017. [Online, accessed 8-June-2026].

\bibitem{uson2021}
J. Arellano-Uson, E. Magana, D. Morato et al., ``Protocol-agnostic method for monitoring interactivity time in remote desktop services,'' Multimed Tools Appl 80, 19107–19135 (2021). https://doi.org/10.1007/s11042-021-10708-3

\bibitem{uson2023}
J. Arellano-Uson, E. Magana, D. Morato and M. Izal, ``Evaluation of RTT as an Estimation of Interactivity Time for QoE Evaluation in Remote Desktop Environments,'' 2023 33rd International Telecommunication Networks and Applications Conference, Melbourne, Australia, 2023, pp. 240-245, doi: 10.1109/ITNAC59571.2023.10368539.




\bibitem{Stockhammer2011acm}
T. Stockhammer, 
``Dynamic adaptive streaming over HTTP,''
in Proc. ACM Conference on Multimedia Systems (MMSys), February 2011.


\bibitem{kimura2021}
T. Kimura, T. Kimura, A. Matsumoto, and K. Yamagishi,
``Balancing Quality of Experience and Traffic Volume in Adaptive Video Streaming,'' \textit{ IEEE Access} 9, pp. 15530 - 15547, 2021.

\bibitem{seufert2015}
M. Seufert, S. Egger, M. Slanina, T. Zinner, T. Ho\ss feld, and P. T.-GiaAuthors, 
``A Survey on Quality of Experience of HTTP Adaptive Streaming,''
\textit{IEEE Communications Surveys \& Tutorials}, 
Vol. 17, issue 1, January 2015.


\bibitem{Sabet2022}
S.~Sabet, S.~Schmid, and A.~El Saddik, ``Quantifying the impact of network delay switching on QoE in online multiplayer games,'' in \emph{Proc. IEEE Global Communications Conference (GLOBECOM 2022)}, Rio de Janeiro, Brazil, Dec.~2022, pp.~3041--3046.


\bibitem{Obafemi2011}
A.~Obafemi, A.~L.~Mohammed, and S.~Misra, ``Impact of jitter playout buffer on E-model in VoIP,'' in \emph{Proc. 10th Int. Conf. on Networks (ICN 2011)}, St.~Maarten, Netherlands Antilles, Jan.~2011, pp.~135--140.




\bibitem{rfc_delayedack}
R. Braden, ``Requirements for Internet Hosts -- Communication Layers'', STD 3, RFC 1122, October 1989.

\bibitem{geoIP}
MAXMIND, ``GeoLite Databases and Web Services,''
\url{https://dev.maxmind.com/geoip/geolite2-free-geolocation-data/}

\bibitem{Nieh2003}
J. A.~Nieh, S. J.~Yang, and N.~Novik,
``Measuring Thin-Client Performance Using Slow-Motion Benchmarking,''
ACM Transactions on Computer Systems, Vol. 21, No. 1, Feb. 2003, pp.~87–-115.


\bibitem{EMA}
A. C. Smit, E. Schat, E. Ceulemans,
``The Exponentially Weighted Moving Average Procedure for Detecting Changes in Intensive Longitudinal Data in Psychological Research in Real-Time: A Tutorial Showcasing Potential Applications,'' Assessment 30, pp. 1354–1368, 2023.


\bibitem{schultz1997}
W.~Schultz, P.~Dayan, and P.~R.~Montague, ``A Neural Substrate of Prediction and Reward,'' \emph{Science}, vol.~275, no.~5306, pp.~1593--1599, Mar. 1997.

\bibitem{lightGBM}
lightgbm.LGBMRegressor, 
\url{https://lightgbm.readthedocs.io/en/latest/pythonapi/lightgbm.LGBMRegressor.html}
[Online; accessed 8-June-2026].

\bibitem{SHAP}
S. Lundberg, and S.-I. Lee,
``A Unified Approach to Interpreting Model Predictions,''
Proceedings of the 31st International Conference on Neural Information Processing Systems (NIPS'17), 
pp. 4768 -- 4777, 2017.





\end{thebibliography}

\begin{acknowledgment}
This research was supported by a grant from the Telecommunications Advancement Foundation.
\end{acknowledgment}

\appendix

%A.1

\section{User Activity and Packet/Byte Count}
\label{A}

To validate that the traffic metrics we observe in the logs are informative proxies for office-work activity levels, rather than, for example, background system traffic, we conducted a controlled traffic measurement experiment in which we ran office applications (Google Docs and Google Slides) via the Thin-Telework System and measured the generated traffic (the numbers of packets sent and received, and the numbers of bytes sent and received). For comparison, we also ran a YouTube video-watching session. For Google Docs and Slides, we prepared two conditions with high and low activity levels. Table~\ref{tab:Context} summarizes the results. We observe that these traffic statistics depend on both the application and the activity level ~\cite{Nieh2003}. Traffic generated by Google Docs and Slides falls within the range observed in our measurement data (Figs.~\ref{fig:Packet_CCDF} and~\ref{fig:Byte_CCDF}). Note that it is biased toward the upper part of the distribution because our experiment did not include user think time and involved only continuous interaction. In contrast, traffic generated by YouTube far exceeds this range.

\begin{table}[tb]
\centering
\caption{Traffic Statistics Per 60 Seconds for Different Applications and Activity Levels}
\label{tab:Context}
\begin{tabular}{lrrrr}
\toprule
User  & \# of Packet & \# of Packet 	& \# of Bytes 	& \# of Bytes \\ 
Activity & Sent	& Received & Sent & Received\\
\midrule
Slide (High) &3,241 &	2,700 &	299 K  	 &1,351 K  \\ \hline
Slide (Low) &2,047 &	1,777 &	184 K 	&734 K  \\ \hline
Docs (High) &2,058 &	3,043 &	176 K 	&2,525 K  \\ \hline
Docs (Low) &1,047 &	1,098 &	92 K& 	359 K \\ \hline
Youtube & 30 K &	77 K &	2,384 K &101,716 K \\ \hline
% inactive 610 	608 	50,977 	45,343 
\bottomrule
\end{tabular}
\end{table}

\begin{biography}
\vspace{-1cm}
\profile{Keisuke ISHIBASHI}{received the B.S. and M.S. degrees in mathematics from Tohoku University, in 1993 and 1995, respectively, and the Ph.D. degree in information science and technology from The University of Tokyo, in 2005. From 1995 to 2018, he was with Nippon Telegraph and Telephone (NTT) Laboratories, where he was involved in research on the measurement and analysis of internet traffic and performance. He is currently a Professor of Information Science with International Christian University, Japan. He is a member of IEICE, IEEE, and the Information Processing Society of Japan (IPSJ).}
\profile{Xuliang DENG}{received the B.A. degree in information science from International Christian University, Japan, in 2023, and the M.S. degree in information networking from Carnegie Mellon University in 2024. He has professional experience as a software engineer in both Japan and the U.S. He is currently a software engineer in industry.}
\profile{Yoshiaki KITAGUCHI}{received the B.S. and M.S. degrees in Physics from Niigata University, Japan in 1995 and 1997, respectively. He joined INTEC Inc. as a Researcher in 1997. He received the Ph.D. degree in Information Systems Engineering from The University of Electro-Communications, Japan in 2005. From 2009 to 2016, he was an Assistant Professor at the Information Media Center, at Kanazawa University, Japan. He is currently an Associate Professor at the Global Scientific Information and Computing Center, Tokyo Institute of Technology, Japan since 2017. He has been engaged in the research and development of IPv6. He is a member of the IEEE Communications Society, ACM, IEICE, and Information Processing Society of Japan (IPSJ).}
\profile{Kenichi NAGAMI}{received the M.S degree in 1992, the Ph.D. in 2001,
both from Tokyo Institute of Technology, Japan.
In 1992, he joined Research and Development Center,
Toshiba Corporation where he focused on communication systems.
Since 2002, he has been with INTEC NetCore, and currently works in
the Research and Development Department at INTEC.}
\profile{Ichiro MIZUKOSHI}{received his B.S. degree in Mathematics from Waseda University, Japan, in 1986, and his M.S. degree in Management Science from Tsukuba University, Japan, in 1992. He has worked on various online services and Internet Service Providers (ISPs). In 1997, he joined NTT (Nippon Telegraph and Telephone Corporation), and since 2006, he has been working at NTT East. He is a member of the Information Processing Society of Japan (IPSJ).}
\profile{Akira SATO}{
received his Ph.D. from University of Tsukuba in 1998. He is an associate professor in Department of Information Engineering, Academic Computing and Communications Center at University of Tsukuba. His current research
interest is an operation of academic networks. He is a member of the Information Processing Society of Japan (IPSJ).
}
\profile{Daiyu NOBORI}{
 is a software engineer and an entrepreneur. His development and research interests include systems software such as Virtual Private Network (VPN), distributed systems, and security. He entered University of Tsukuba in 2003 and started up a company, SoftEther Corporation in 2004. He acquired a Ph.D. degree at the Department of Computer Science, University of Tsukuba in 2017. He has been an visiting professor at University of Tsukuba since April 2022. He has developed SoftEther VPN, a cross-platform multi-protocol VPN program, made it public for free, and opened its source code in 2014.
}
\end{biography}

\end{document}